\documentclass[10pt,twocolumn,english,aps,prb]{revtex4}
\usepackage[latin1]{inputenc}
\usepackage{babel}
\usepackage{graphics}
\usepackage{amssymb}

\makeatletter

\providecommand{\LyX}{L\kern-.1667em\lower.25em\hbox{Y}\kern-.125emX\@}

\makeatletter
\date{}

\newcommand{\ket}[1]{{| #1 \rangle}}

\newcommand{\then}{\Rightarrow}

\newcommand{\bbbn}{{\rm I\!N}}

\newcommand{\real}{ {\rm I} \hspace{-1.0mm} {\rm Re} \: }
\newcommand{\imag}{ {\rm I} \hspace{-1.0mm} {\rm Im} \:}

\makeatother
\begin{document}

\title{Fermi liquid parameters in 2D with spin-orbit interaction}

\author{D. S. Saraga and Daniel Loss}

\address{{\footnotesize Department of Physics and Astronomy, University of
Basel, Klingelbergstrasse 82, CH-4056 Basel, Switzerland }}

\date{\today }

\begin{abstract}
We derive analytical expressions for the quasiparticle lifetime \( \tau  \),
the effective mass \( m^{*} \), and the Green's function renormalization
factor \( Z \) for a 2D Fermi liquid with electron-electron interaction
in the presence of the Rashba spin-orbit interaction. 
We find that the modifications
are independent of the Rashba band index \( \rho  \), and occur in
second order of the spin-orbit coupling \( \alpha  \).
In the derivation of these results, we also discuss
the screening of the Coulomb interaction, as well as the susceptibility and the
self-energy in small $\alpha$.
%\\
\\
PACS: 71.10.Ca, 73.21.-b, 71.18.+y
\end{abstract}
\maketitle

\section{Introduction}

The lifetime of quasiparticle excitations determined by electron-electron
collisions is a crucial quantity of the Fermi liquid theory \cite{Mah00}
of interacting electron systems. In particular, the quasi-particle
lifetime \( \tau  \) for a two-dimensional electron gas as found
e.g. in semiconductor heterostructures has been now studied in great
detail \cite{Giu82,Zhe96}. While \( \tau  \) has been traditionally
of importance for phenomena that are based on coherent transport such
as for example conductance fluctuations, weak-localization, or the
Aharonov-Bohm effect \cite{Was92}, this quantity is also important
for the current strive towards quantum information processing in the
solid state, which requires the coherent propagation of e.g. entangled
electrons. In this respect, as well as in the emerging field of spintronics,
the spin degree of freedom is increasingly being investigated \cite{ALS02}. 

The effect of spin-orbit (s-o) interaction in low-dimensional systems
has consequently become an important issue, and has uncovered new
functionalities such as the spin-based transistor \cite{Dat90}, spin
injection \cite{Vos00} and the electric manipulation of spin in non-magnetic
semiconductors \cite{Kat04}, and has also led to new physics with
the spin-Hall effect$^{9-15}$.
%\cite{Mur03,Sin03,Kat04hall,Wun05,Sch04,Cha04,Erl05}.
The consideration of s-o interaction in the framework of Fermi liquid
theory is therefore desireable. Existing work has investigated electronic
transport and plasmon excitations \cite{Mis03,Mag01}, Friedel-like
oscillations in the screened potential \cite{Che99}, and the modification
of the s-o coupling due to electron-electron interactions\cite{Che99b}.
While the spin relaxation and decoherence rates have been widely studied
in such systems\cite{Ell54}, the relaxation rate of the quasiparticle
itself has not, to our knowledge, been studied so far. 

An important contribution to the effective mass \( m^{*} \) comes
from the renormalization of the electron band mass by electron-electron
interactions. Simple expressions for \( m^{*} \) in 2D appear in
early works addressing the \emph{\( g \)-}factor \cite{Jan69} and
the spin susceptibility \cite{Yar89}, and 
were followed by numerical studies \cite{Tin75}. % while 
Some recent
work addressed non-analytic corrections\cite{Chu03}, the temperature
dependence \cite{Zha04}, and the effects of impurity scattering \cite{Mor04}.
Another important parameter of Fermi liquid theory is the renormalization
factor \( Z \) of the Green's function \cite{Mah00}. This quantity
measures the quasiparticle spectral weight, and gives the size of
discontinuity of the zero temperature Fermi occupation factor \( n(\xi ) \)
at the Fermi surface. For a clean 2DEG without impurities and s-o
interaction, it has first been studied for short-range potentials
\cite{Blo75}, while the realistic case with Coulomb interaction has
been studied numerically \cite{Jal89} and analytically \cite{Bur00,Gal04}.  Recent related
work used Fermi liquid theory to study plasmons contributions to the effective mass in
valley-degenerate systems \cite{Gan05}, spin resonance and the spin-Hall 
conductivity \cite{She05}, as well as screening and plasmon modes \cite{Ple05}

This work presents an analytical study of the effect of s-o interaction
on the quasiparticle lifetime, the \( Z \)-factor, and the effective
mass \( m^{*} \) in a two dimensional Fermi liquid, taking the specific
case of the Rashba interaction \cite{Byc84}. We consider the long-range
Coulomb interaction, and work within the random phase approximation\cite{Mah00}
(RPA) valid for small \( r_{s}\ll 1 \) (high densities). For the
lifetime, we find that the spin-orbit contribution appears in second
order of the s-o coupling \( \alpha  \), and contains a logarithmic
term similar to the standard lifetime \cite{Giu82}, where the excitation
energy \( \xi  \) is replaced by the Rashba splitting \( 2\alpha k_{F}/\hbar  \).
A similar result is found for the effective mass, with a modification
of the form $ \alpha^2 \log\alpha $. 
For the $Z$-factor, we find a quadratic term without logarithmic enhancement.
In all these cases the modifications
are independent of the Rashba band index \( \rho  \) denoting the
two directions of the eigenspinors of the Rashba Hamiltonian.
We also discuss briefly the screening of the Coulomb interaction, 
and derive expressions
both the real and imaginary parts of the susceptibility \( \chi  \),
complementing the expressions found in Refs. \cite{Che99,Mis03,Mag01}.
We also give general arguments showing that the self-energy and, consequently,
the Fermi liquid parameters, cannot have any modification linear in $\alpha$.
Throughout this work we consider a clean system at zero temperature.
%\nopagebreak
\section{2D Fermi liquid with Rashba s-o interaction}

\subsection{Rashba eigenstates}

We consider an electron in a 2D Fermi liquid in the presence of the
Rashba spin-orbit interaction restricted to the \( z=0 \) plane,
described by the Hamiltonian \( H=p^{2}/2m+H_{\mathrm{so}} \) with
\cite{Byc84}\begin{equation}
\label{1}
H_{\mathrm{so}}=\frac{\alpha }{\hbar }\left( p_{x}\sigma ^{y}-p_{y}\sigma ^{x}\right) .
\end{equation}
Expressed in the \( \sigma _{z} \)- spin basis \( \ket {\pm }_{z} \),
the eigenstates are \begin{equation}
\label{eigen}
\ket {\mathbf{k},\rho }=\frac{1}{\sqrt{2}}\left( \begin{array}{c}
1\\
i\rho e^{i\phi (\mathbf{k})}
\end{array}\right) \ket {\mathbf{k}},
\end{equation}
with the polar angle \( \phi (\mathbf{k})=\angle (\mathbf{k},\mathbf{Ox}) \)
and the momentum eigenstates \( \ket {\mathbf{k}} \). The index \( \rho =\pm  \)
defines the two Rashba bands. We define the s-o strength \begin{equation}
\label{1}
\gamma =\frac{k_{R}}{k_{F}}=\frac{\alpha k_{F}/\hbar }{2E_{F}}=\sqrt{\frac{E_{R}}{2E_{F}}}
\end{equation}
 from the Rashba momentum\begin{equation}
\label{1}
k_{R}=m\alpha /\hbar ,
\end{equation}
 the Rashba energy \( E_{R}=m\alpha ^{2}/\hbar ^{2} \), and the Fermi
energy \( E_{F}=k_{F}^{2}/2m \). We define the excitation energy
by \( \xi _{k\rho }=E_{k\rho }-E_{F} \), with the dispersion relations
for the two branches \( E_{k\rho }=\left( k^{2}+2\rho k_{R}k\right) /2m \).
Setting \( \xi _{k_{\rho }\rho }=0 \) yields the two Fermi momenta
\begin{equation}
\label{1}
k_{\rho }=\kappa -\rho k_{R}\, \, \, \, \mathrm{with}\, \, \, \, \kappa =k_{F}\sqrt{1+\gamma ^{2}}.
\end{equation}
Note that both \( k_{\rho } \) and \( \kappa  \) will replace \( k_{F} \)
in a number of the expressions valid without s-o interaction . We
define the unperturbed Matsubara \cite{Mah00} Green's function \begin{equation}
\label{unpert-green}
G_{\rho }(k,ik_{n})=\frac{1}{ik_{n}-\xi _{k\rho }}
\end{equation}
corresponding to the Rashba eigenstates (\ref{eigen}) without electron-electron
interaction. We have introduced the fermionic Matsubara frequency
\( k_{n}=(2n+1)\pi k_{B}T,n\in \bbbn  \).

\subsection{Renormalization due to the electron-electron interaction.}

Within Fermi liquid theory, the presence of electron-electron interaction
modifies the retarded Green's function \cite{Mah00} \begin{eqnarray}
\bar{G}^{R}_{\rho }(k,\xi ) & = & \bar{G}_{\rho }(k,ik_{n}\to \xi +i0^{+})=\frac{1}{\xi -\xi _{k\rho }-\Sigma ^{R}_{\rho }(k,\xi )}\nonumber \\
 & \simeq  & \frac{Z_{\rho }}{\xi -\xi ^{*}_{k\rho }+i(\hbar /2)\Gamma _{\rho }(k)}\label{1} 
\end{eqnarray}
describing a quasiparticle belonging to the Rashba band \( \rho  \)
with a momentum \( \mathbf{k} \). To derive the expression above,
one has expanded for small frequencies \( \xi  \) and small excitation
energies \( \xi _{k\rho } \) above the Fermi surface, i.e. \( \xi \ll E_{F} \),
\( 0<\xi _{k\rho }\ll E_{F}\, \Leftrightarrow  \) \( k-k_{\rho }\ll k_{\rho } \).
In this procedure, one first shifts the Fermi momentum \( k_{\rho } \)
via the requirement \( \xi _{k_{\rho }\rho }+\real \Sigma ^{R}_{\rho }(k_{\rho },0)=0 \).
The lifetime of the quasiparticle, \( \tau _{\rho }(k)=1/\Gamma _{\rho }(k) \),
is given via \begin{equation}
\label{life}
\Gamma _{\rho }(k)=-\frac{2}{\hbar }\imag \Sigma ^{R}_{\rho }(k,\xi _{k\rho }),
\end{equation}
where the self-energy \( \Sigma  \) contains the effect of the Coulomb
electron-electron interaction. 

The Green's function acquires a renormalized weight\begin{eqnarray}
 &  & Z_{\rho }=\frac{1}{1-A},\label{zfactor} \\
\mathrm{with} &  & \, A:=\frac{\partial }{\partial \xi }\real \Sigma ^{R}_{\rho }(k_{\rho },\xi =0)\label{zfactor-A} 
\end{eqnarray}
 which gives the size of the jump in the Fermi occupation factor \( n(\xi ) \)
at the Fermi surface. 

The effective mass enters in the renormalized excitation energy \( \xi ^{*}_{k\rho }=\left( k^{2}+2\rho k_{R}k-k_{F}^{2}\right) /2m^{*} \),
and is defined by\begin{eqnarray}
 &  & \frac{m^{*}_{\rho }}{m}=\frac{1}{Z_{\rho }}\frac{1}{1+B},\label{mass} \\
\mathrm{with} &  & B:=\frac{m}{\kappa }\frac{\partial }{\partial k}\real \Sigma ^{R}_{\rho }(k=k_{\rho },0).\label{mass-B} 
\end{eqnarray}
As the excitation energy must vanish at the Fermi surface, one has
\( \xi ^{*}_{k_{\rho }\rho }=0 \) and thus \( k_{F} \) is also shifted
with \( k_{\rho } \). Note that it is \( \kappa  \), not \( k_{\rho } \),
that enters in the factor \( m/\kappa  \) appearing in \( B \). 

In order to study the modifications introduced by the Rashba interaction,
we first present here the results found without s-o interaction. The
inverse lifetime reads\cite{Giu82,Zhe96,Lifediffer} \begin{equation}
\label{G0}
\Gamma _{0}(k)=\frac{\xi _{k}^{2}}{E_{F}}\left[ \log \frac{\xi _{k}}{E_{F}}+O\left( r_{s}\right) \right] .
\end{equation}
The \( r_{s} \)-factor is defined here \cite{rs} as \( r_{s}=k_{\mathrm{TF}}/2k_{F}=m e_{0}^{2}/\hbar ^{2}\sqrt{2\pi n} \),
where \( n \) is the electronic sheet density (in the absence of s-o interaction), and \( k_{\mathrm{TF}} \)
is the Thomas-Fermi screening momentum. The two important characteristics
of (\ref{G0}) are: (i) \( \Gamma _{0}\to 0 \) when \( \xi _{k}\to 0 \),
corresponding to long-lived quasiparticle excitations near the Fermi
surface; (ii) the vanishing of \( \Gamma  \) is slowed down by a
logarithmic factor. 

The effective mass contains a term \( \sim r_{s}\log r_{s} \), and
is given by \cite{Jan69} \begin{equation}
\label{meff0}
\frac{m_{0}^{*}}{m}-1=\frac{r_{s}}{\pi }\left[ \log r_{s}+2-\log 2+O(r_{s})\right] .
\end{equation}

The deviation of the renormalization weight \( Z \) from \( 1 \)
is linear with \( r_{s} \), and reads \cite{Bur00}\begin{equation}
\label{z0}
Z_{0}-1=-\frac{r_{s}}{\pi }\left[ 1+\frac{\pi }{2}+O(r_{s})\right] .
\end{equation}
Taking a GaAs 2DEG with \cite{And82} \( n=4\cdot 10^{15}\, \mathrm{m}^{-2} \),
one has \( r_{s}=0.614 \), \( Z_{0}=0.50 \), and \( m^{*}_{0}/m-1=0.16 \).
An InAs 2DEG with e.g. \cite{Sat01} \( n=10\cdot 10^{15}\, \mathrm{m}^{-2} \),
\( m=0.03m_{e} \) and \( r_{s}=0.18 \) has the parameters \( Z_{0}=0.83 \)
and \( m^{*}_{0}/m-1=0.019 \).

\subsection{Screening of the Coulomb interaction}

\begin{figure}[t]
{\centering \resizebox*{0.98\columnwidth}{!}{\includegraphics{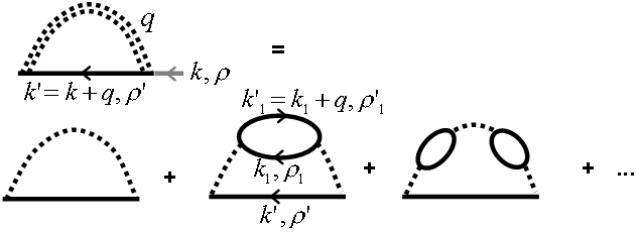}} \par}

\caption{\label{diagram}The diagramatic representation of the self-energy
\protect\( \Sigma _{\rho }(k)\protect \) (\ref{selfener}) in RPA.
The full lines denote the electron Green's functions (\ref{unpert-green}),
the dashed lines the Coulomb interaction, the circles are the susceptibility
bubble diagram (\ref{bubble}), and the double dashed line is the
screened Coulomb interaction (\ref{screened-coulomb}). The Rashba
bands are denoted by the \protect\( \rho \protect \)'s, and yield
the overlap factor \protect\( \mathcal{F}'\protect \) (\ref{fprime}).}
\end{figure}
In order to build a Fermi liquid theory including the s-o interaction,
we must consider the matrix elements of the bare 2D Coulomb interaction
\( V_{C}(q)=2\pi e_{0}^{2}/q\hbar  \) in the Rashba eigenstates basis.
These matrix elements involve the overlap \begin{equation}
\label{1}
\mathcal{F}=\frac{1}{4}\left[ 1+\rho _{1}\rho '_{1}e^{i(\phi _{1}'-\phi _{1})}\right] \left[ 1+\rho _{2}\rho '_{2}e^{i(\phi _{2}'-\phi _{2})}\right] 
\end{equation}
 of the eigenspinors (\ref{eigen}), which depends on the directions
\( \{\phi _{1},\phi _{2},\phi _{1}',\phi _{2}'\} \) of the scattered
states and on their band indices \( \{\rho _{1},\rho _{2},\rho _{1}',\rho _{2}'\} \).
In RPA\cite{Mah00}, we find the screened Coulomb potential\begin{equation}
\label{screened-coulomb}
V(q,\omega )=\frac{V_{C}(q)}{\varepsilon (q,\omega )}\mathcal{F},
\end{equation}
 where \( \varepsilon =1-V_{C}\chi  \) is the dielectric function.
In Matsubara formalism \cite{Mah00}, the susceptibility is given
by the bubble diagram\begin{equation}
\label{bubble}
\chi (\mathbf{q},iq_{n})=k_{B}T\sum _{\rho ,\rho '}\sum _{\mathbf{k},ik_{n}}G_{\rho }(\mathbf{k},ik_{n})G_{\rho '}(\mathbf{k}',ik'_{n})\mathcal{F}',
\end{equation}
 where \( \mathbf{k}'=\mathbf{k}+\mathbf{q} \), \( ik_{n}'=ik_{n}+iq_{n} \),
\( \theta =\angle (\mathbf{k},\mathbf{k}') \), \( (2\pi )^{2}\sum _{\mathbf{k}}\to \int d\mathbf{k} \)
and \( q_{n}=2n\pi k_{\mathrm{B}}T \) is a bosonic Matsubara frequency.
The corresponding diagrams are represented in Fig. \ref{diagram}.
The last factor\begin{equation}
\label{fprime}
\mathcal{F}'=\frac{1+\rho \rho '\cos \theta }{2}
\end{equation}
 is the overlap \( \mathcal{F} \) for states with opposite momenta.
After summing over \( ik_{n} \) and performing the analytical continuation
\( iq_{n}\to \omega +i0^{+} \), one finds \begin{equation}
\label{rechir}
\real \chi (\mathbf{q},\omega )=\sum _{\mathbf{k},\rho \rho '}\frac{n\left( \xi _{k\rho }\right) -n\left( \xi _{k'\rho '}\right) }{\omega +\xi _{k\rho }-\xi _{k'\rho '}}\mathcal{F}',
\end{equation}
\begin{eqnarray}
\imag \chi (\mathbf{q},\omega ) & = & -\pi \sum _{\mathbf{k},\rho \rho '}\delta \left( \omega +\xi _{k\rho }-\xi _{k'\rho '}\right) \nonumber \\
 &  & \hspace {-0mm}\times \, \left[ n\left( \xi _{k\rho }\right) -n\left( \xi _{k'\rho '}\right) \right] \mathcal{F}',\label{imchir} 
\end{eqnarray}
which is the standard form with the additional \( \mathcal{F}' \)
factors. As we consider zero temperature, the Fermi occupation factor
reads \( n(\xi )=\Theta (-\xi ) \). Note that the effect of the spin-orbit
interaction manifests itself in the energies \( \xi _{\rho k} \),
while the factors \( \mathcal{F}' \) \emph{alone} just describe a
change of the spin basis. In particular, such a basis change could
also be considered in the absence of s-o interaction. For instance,
the scattering cross-section for two electrons \cite{Sar04}, given
in Born approximation by \begin{equation}
\label{1}
\lambda =\frac{1}{2\pi k}\left| \frac{m}{\hbar ^{2}}V(q)\right| ^{2},
\end{equation}
 vanishes for different Rashba bands in case of forward scattering
(e.g \( \rho _{1}'=-\rho _{1} \), \( \mathbf{k}_{1}'=\mathbf{k}_{1} \)),
while it vanishes for same bands in the case of backscattering (\( \mathbf{k}_{1}'=-\mathbf{k}_{1} \)).
This only reflects the fact that the real spin is conserved by the
Coulomb interaction. For forward scattering the spin basis does not
change, so that the band index must be the same. The opposite happens
for backscattering, where the spin basis is inverted and the band
index must be changed in order to preserve the real spin. This is
the same reason why the two conjugate states of the Kramers doublet
belong to the same Rashba band \cite{Ras03}.

\subsection{Self-energy}

The self-energy is the central quantity that determines the other
Fermi-liquid parameters. In lowest order in the screened intraction
(RPA),  it is given by \cite{Mah00}\begin{equation}
\label{selfener}
\Sigma _{\rho }(k,ik_{n})=-k_{B}T\sum _{\rho '}\sum _{\mathbf{q},iq_{n}}G_{\rho '}(\mathbf{k}',ik'_{n})V'(q,iq_{n}),
\end{equation}
and is represented in Fig. \ref{diagram}. Here the screened potential
(\ref{screened-coulomb}) \( V'=\mathcal{F}'V_{C}/\varepsilon  \)
involves \( \mathcal{F}' \) because of momentum conservation. At
zero temperature the inverse lifetime \( 1/\tau _{\rho }=\Gamma _{\rho } \)
is given by\begin{equation}
\label{inverselife2}
\Gamma _{\rho }(k,\xi _{k\rho })=\frac{2}{\hbar }\sum _{\mathbf{q}\rho '}\Theta (\xi _{k'\rho '})\Theta (\omega )\, \imag V'(q,\omega ),
\end{equation}
with \( \omega =\xi _{k\rho }-\xi _{k'\rho '} \).We now introduce
the parameters \begin{equation}
\label{1}
x''=\frac{q}{2k_{\rho }}\, \, \mathrm{and}\, \, y'=\frac{m\omega }{q\kappa },
\end{equation}
which are relevant for the susceptibility \( \chi  \) entering in
\( V \) (see Sec. \ref{sec-pol}). We consider small excitation energies
above the Fermi surface\begin{equation}
\label{1}
0<\xi _{k\rho }\simeq \frac{\kappa \, \delta }{m}\ll E_{F}\, \, \Leftrightarrow \, \, \delta :=k-k_{\rho }\ll k_{\rho }.
\end{equation}
 Using \( \omega =\left[ k^{2}-k'^{2}+2k_{R}(\rho k-\rho 'k')\right] /2m \),
we see that the \( \Theta  \) functions in (\ref{inverselife2})
yield \begin{equation}
\label{1}
0<y'<\bar{y}'\sim \frac{\mathrm{max}\{\delta ,k_{R}\}}{q}.
\end{equation}
Note that a priori, neither \( x'' \) nor \( y' \) have to be small;
however, one can check numerically that the dominant contributions
to (\ref{inverselife2}) come from forward scattering, i.e. for\begin{equation}
\label{1}
q\lesssim \delta \ll k_{\rho }\, \, \Leftrightarrow \, \, x''\ll 1.
\end{equation}
For the lifetime, one can also assume \( y'\ll 1 \). For these reasons
we shall calculate the susceptibility in the limit \( x''\ll 1 \),
before taking \( y'\ll 1 \).

\section{susceptibility\label{sec-pol}}

The susceptibility \( \chi  \) (or, equivalently, the dielectric
function \( \epsilon =1-V_{C}\chi  \)) for a 2DEG with s-o interaction
has been partially studied in Refs.\cite{Mis03,Mag01,Che99} in the
small \( q\ll k_{F} \) limit. Expressions for the imaginary part
of \( \chi  \) in the limit \( q\ll k_{R},k_{F} \) have been given
in Refs. \cite{Mis03,Mag01} in the context of transport. Ref. \cite{Che99},
which addressed non-analytic contributions to the real part of \( \epsilon  \),
only gives expressions for \( q\to 0 \) for the interband case (different
Rashba bands), while the intraband case is studied in the \( q\to 2k_{F} \)
case. Therefore, it is desireable to complement these studies by deriving
expressions for both \( \real \chi  \) and \( \imag \chi  \) in
the \( q\to 0 \) limit.

We first write the 2D susceptibility \( \chi _{0} \) without Rashba
interaction \cite{Ste67}. Introducing the parameters\begin{equation}
\label{1}
x=\frac{q}{2k_{F}}\, \, \mathrm{and}\, \, y=\frac{m\omega }{qk_{F}},
\end{equation}
 the susceptibility reads in the Matsubara formalism (\( iy_{n}\to y+i0^{+} \))
\begin{equation}
\label{chi matsubara}
\chi _{0}(x,iy_{n})=-\frac{m}{\pi }\left\{ 1-\frac{1}{x}R\left[ s(z)\sqrt{z^{2}-1}\right] \right\} ,
\end{equation}
where \( z=x+iy_{n} \), \( z^{*}=x-iy_{n} \), \( R\left[ f\left( z\right) \right] =\left[ f(z)+f(z^{*})\right] /2 \),
and \( s(z)=\mathrm{sgn}\left[ \real (z)\right]  \) arising from
the choice of the \( (-\infty ,0] \) branch cut for \( \sqrt{z} \). 

We now derive the susceptibility (\ref{rechir})-(\ref{imchir}) in
the limit of small \( q,k_{R}\ll k_{\rho } \). We first write \( \chi (q,\omega )=\sum _{\rho \rho '}\chi _{\rho ,\rho '} \).
We define \( \varphi =\angle (\mathbf{k},\mathbf{q}) \), and expand
\( k'=k\sqrt{1+2\cos \varphi \, (q/k)+(q/k)^{2}} \) in small \( q \)
to get the energy difference\begin{equation}
\label{1}
\xi _{k\rho }-\xi _{k'\rho '}\simeq \frac{k}{m}k_{R}(\rho -\rho ')-\frac{q}{m}(k+\rho 'k_{R})\cos \varphi .
\end{equation}
We also expand the Fermi function \begin{equation}
\label{1}
n(\xi _{k'\rho '})-n(\xi _{k\rho })\simeq \left( \xi _{k\rho }-\xi _{k'\rho '}\right) \delta (\xi _{k\rho }),
\end{equation}
 which selects \( k=k_{\rho } \), and the spinor overlap\begin{equation}
\label{1}
\frac{1\pm \cos \theta }{2}=\frac{1}{2}\left( 1\pm \frac{k+q\cos \varphi }{k'}\right) \simeq \frac{1}{2}\pm \left[ \frac{1}{2}-\left( \frac{q}{2k}\sin \varphi \right) ^{2}\right] .
\end{equation}
These expansions are valid for \( q,k_{R}\ll k_{\rho } \).

\subsection{Intraband contributions (\protect\( \rho '=\rho \protect \))}

We first consider transitions within a given Rashba band. We can neglect
\( (q/k)^{2} \) in the spinor overlap, and integrate over \( \varphi  \)
and \( k \). We find \begin{equation}
\label{intrare}
\real \chi _{\rho ,\rho }=-\frac{m}{2\pi }\left[ 1-\frac{|y'|}{\sqrt{y'^{2}-1}}\Theta (|y'|-1)\right] \left( 1-\rho \frac{k_{R}}{\kappa }\right) .
\end{equation}
For the imaginary part, \( \delta (\omega +\xi _{k\rho }-\xi _{k'\rho '}) \)
selects \( \varphi =\mathrm{Arccos}(y') \) if \( |y'|<1 \), and
we get \begin{equation}
\label{intraim}
\imag \chi _{\rho ,\rho }=-\frac{m}{2\pi }\frac{y'}{\sqrt{1-y'^{2}}}\Theta (1-|y'|)\left( 1-\rho \frac{k_{R}}{\kappa }\right) ,
\end{equation}
which agrees with Eq. (35) of Ref. \cite{Mis03}. Summing over \( \rho  \),
we see that the intraband contributions to \( \chi  \) are independent
of the band index \( \rho  \).

\subsection{Interband contributions (\protect\( \rho '=-\rho \protect \))}

For transitions between two different Rashba bands, it is necessary
to distinguish between two cases. 

(a) \( k_{R}\ll q\ll k_{\rho } \). We find\begin{equation}
\label{intere1}
%\real \chi _{\rho ,-\rho }=-\frac{m}{2\pi }\frac{1}{2}x^{2}\left[ 1+\left( |y|\sqrt{y^{2}-1}-y^{2}\right) \Theta (|y|-1)\right] ,
\real \chi _{\rho ,-\rho }=-\frac{m}{2\pi} x^{2}\left[ \frac{1}{2}+\left( |y|\sqrt{y^{2}-1}-y^{2}\right) \Theta (|y|-1)\right] ,
\end{equation}
\begin{equation}
\label{interim1}
\imag \chi _{\rho ,-\rho }=-\frac{m}{2\pi }x^{2}y\sqrt{1-y^{2}}\Theta (1-|y|),
\end{equation}
where we have also expanded in \( y'\simeq y \).

(b) \( q\ll k_{R},k_{\rho } \). We get\begin{eqnarray}
\real \chi _{\rho ,-\rho } & = & \frac{1}{16\pi }\frac{q^{2}}{\omega }\log \left( \frac{\rho \kappa /k_{R}+4E_{F}/\omega -1}{\rho \kappa /k_{R}+4E_{F}/\omega +1}\right) \nonumber \\
 & \simeq  & -\frac{m}{4\pi }x^{2}\frac{1}{1+m\omega /2\rho k_{R}k_{F}},\label{intere2} 
\end{eqnarray}
where we have expanded in small \( k_{R}\ll \kappa  \) in the second
equality. Note the unusal term \( m\omega /k_{R}k_{F} \). Setting
\( \omega =0 \) and summing over \( \rho  \) yields the static result
(24) of Ref. \cite{Che99} in the limit \( k_{B}T\to 0 \). For the
imaginary part we find \begin{equation}
\label{interim2}
\sum _{\rho }\imag \chi _{\rho ,-\rho }=-\frac{m}{8}\, \frac{x}{y}\Theta \left( \omega _{-}<|\omega |<\omega _{+}\right) ,
\end{equation}
with \( \omega _{\pm }=2(\kappa \pm k_{R})k_{R}/m \). This expression
(\ref{interim2}) agrees with Eq. (37) of Ref. \cite{Mis03} and Eq.
(10) of Ref. \cite{Mag01}, which are relevant for the optical conductivity.
One can neglect this contribution when calculating the lifetime, as
in this case \( q\sim \delta ,\, \omega \sim k_{R}k_{F}/m \) \( \then  \)
\( x/y\ll y \). The other interband contributions are negligible
compared to the intrabands ones, as they are smaller by the factor
\( x^{2}\ll 1 \).

\subsection{Total susceptibility \protect\( \chi \protect \)}

Adding the two intrabands branches, we find for the susceptibility\begin{equation}
\label{1}
\real \chi (q\to 0,\omega )=-\frac{m}{\pi }\left[ 1-\frac{|y'|}{\sqrt{y'^{2}-1}}\Theta (|y'|-1)\right] 
\end{equation}
\begin{equation}
\label{1}
\imag \chi (q\to 0,\omega )=-\frac{m}{\pi }\frac{y'}{\sqrt{1-y'^{2}}}\Theta (1-|y'|)
\end{equation}
which corresponds to the case without Rashba interaction (\ref{chi matsubara})
in the limit \( x\to 0 \), where one replaces \( y \) by the new
parameter \( y' \). We can now take the limit of small energy, \( y'\ll 1 \),
and we finally find the susceptibility in the presence of Rashba s-o
interaction\begin{equation}
\label{chifin}
\chi (q\to 0,\omega \to 0)=-\frac{m}{\pi }\left( 1+iy'\right) ,
\end{equation}
which we shall use in the calculations of the Fermi-liquid parameters below. 
Note that in
general Eqs. (\ref{rechir}) and (\ref{imchir}) yields that $\chi (q,-\omega)=\chi^*(-q,\omega)$.
In particular, $\real \chi $ and
$\imag \chi$ are respectively even and odd in $y$ in the limit $q=0$, as seen in our expressions above.

\section{Susceptibility and self-energy for small $\alpha$}

In this section we show on general grounds that the expansion of \( \chi  \) and $\Sigma$
in small \( \alpha  \) have no term linear in $\alpha$.

\subsection{Susceptibility}
The susceptibility has only a second-order
contribution from the s-o interaction, \begin{equation}
\label{1}
\chi =\chi _{0}+O(\alpha^{2}),
\end{equation}
because  \( \chi  \) is an even function of \( k_R=m \alpha/\hbar \). 
This can be seen by expanding $\chi$ around $\alpha=0$ via the function
$ h(\xi _{k\rho };\xi _{k'\rho '})
:=\left[ n\left( \xi _{k\rho }\right) -n\left( \xi _{k'\rho '}\right) \right] /\left( i\omega _{n}+\xi _{k\rho }-\xi _{k'\rho '}\right)$. 
We use \( d\xi _{k\rho }/dk_{R}=\rho k/m \)
and \( \xi _{k}=k^{2}/2m-E_{F} \), and find
\begin{eqnarray}
\chi (q,i\omega _{n}) & = & \frac{1}{2}\sum _{\rho \rho '}\sum _{\mathbf{k}}\left( 1+\rho \rho '\cos \theta \right) \nonumber \\
 &  & \hspace {-20mm}\times \sum _{j\ge 0}\frac{1}{j!}\left( \frac{k_{R}}{m}\right) ^{j}\left( \rho k\frac{\partial }{\partial \xi _{k}}+\rho 'k'\frac{\partial }{\partial \xi _{k'}}\right) ^{j}h(\xi _{k};\xi _{k'}),\label{1} 
\end{eqnarray}
and notice that the sums over \( \rho ,\rho ' \) cancel for odd
powers \( j \). 
This result is consistent with Eqs. (\ref{intrare})-(\ref{interim2}),
where the terms linear in \( k_{R} \) vanish after summing over \( \rho ,\rho ' \).

\subsection{Self-energy} \label{seflnolin}
We first perform the sum over the Rashba band index $\rho'$ in the self-energy (\ref{selfener})
\begin{equation}
\label{H}
\Sigma _{\rho }(k,ik_n)=-k_{B}T\sum _{\mathbf{q},iq_{n}} H(\mathbf{q},ik'_n)
\frac{V_C(q)}{\varepsilon(q,iq_n)},
\end{equation}
where 
\begin{equation}
 H(\mathbf{q},ik'_n)=
\frac{\zeta+ \rho \alpha (k+q \cos\phi)}
{\zeta^2-(\alpha k')^2},
\end{equation}
and $\zeta=i k'_n -\xi_{k'}$. 
We expand in small $\alpha$ and find
\begin{equation}
 H(\mathbf{q},ik'_n)\simeq
\frac{1}{\zeta_0} +O(\alpha^2)\left( \frac{1}{\zeta_0^2}+\frac{1}{\zeta_0^3}\right) .
\end{equation}
where $\zeta_0=\zeta(\alpha\to 0)$.
(We recall that $k$, being close to $k_\rho$, depends on $\alpha$.)

The integrations of the first and third terms do not yield any logarithmic term in
$\alpha$ because their divergence at $\zeta_0=0$ is odd with respect to $q$.
On the contrary, the term $O(\alpha^2)/\zeta_0^2$ brings logarithmic contributions, 
as will be seen in the lifetime and the effective mass below.
Because $\varepsilon=1-V_C \chi$ has also no term linear in $\alpha$, 
we find that the modification of the self-energy due to the s-o interaction can
only appear in second order.

\section{Lifetime}

In this section we calculate the lifetime as given by Eq.(\ref{life}).
We first define the Thomas-Fermi momentum \begin{equation}
\label{1}
k_{\mathrm{TF}}=\frac{2me_{0}^{2}}{\hbar }=2r_{s}k_{F},
\end{equation}
 and assume  that the small \( q \) contributions dominate such
that \( mV(q)/\pi =k_{\mathrm{TF}}/q\gg 1 \) (this is justified in
GaAs where \( k_{s}\simeq 1.2k_{F} \)). We find\begin{eqnarray}
\imag V'(q,\omega ) & = & \frac{V_{C}^{2}(q)}{(1+k_{\mathrm{TF}}/q)^{2}+(k_{\mathrm{TF}}/q)^{2}y'^{2}}\left( -\frac{m}{\pi }y'\right) \mathcal{F}'\nonumber \\
 & \simeq  & -\frac{\pi }{m}y'\mathcal{F}'.\label{1} 
\end{eqnarray}
Note that it is \( \mathcal{F}' \) --and not \( \mathcal{F}'^{2} \)--
that appears here with \( V^{2} \), because the screening involves
only \( \chi V \), without \( \mathcal{F}' \). Writing \( \Gamma _{\rho }(k)=\sum _{\rho '}\Gamma _{\rho ,\rho '}(k) \)
and changing variables \( \sum _{\mathbf{q}}\to \sum _{\mathbf{k}'} \),
we have\begin{equation}
\label{1}
\Gamma _{\rho ,\rho '}(k)=\frac{1}{8\pi \hbar m\kappa }\, \int _{k_{\rho '}}^{\bar{k}}dk'\, k'\, (\xi _{k'\rho '}-\xi _{k\rho })I_{\rho '},
\end{equation}
where \begin{equation}
\label{1}
I_{\rho '}=\int _{0}^{2\pi }d\theta \frac{1+\rho \rho '\cos \theta }{q(k',\theta )}.
\end{equation}
Here \( \bar{k}=k+k_{R}(\rho -\rho ') \), \( \theta =\angle (\mathbf{k},\mathbf{k}') \),
and \( q(k',\theta )=\sqrt{k^{2}+k'^{2}-2kk'\cos \theta } \). We
distinguish intra- and interband contributions.

\subsection{Intraband case (\protect\( \rho '=\rho \protect \)). }

We find\begin{equation}
\label{1}
I_{\rho }=\frac{2}{kk'|k-k'|}\left[ (k+k')^{2}K(-z)-(k-k')^{2}E(-z)\right] ,
\end{equation}
where \( z=4kk'/(k-k')^{2} \), and \( K \) and \( E \) are the
complete Elliptic integrals of the first and second kind, respectively.
We use their asymptotics \cite{Abr65} \( E(-z)\sim \sqrt{z} \) and
\( K(-z)\sim \log \left( 4\sqrt{z}\right) /\sqrt{z} \) for large
\( z\gg 1 \), as \( k-k'\sim \delta \ll k\simeq k_{\rho } \). After
performing the \( k \) integration and expanding in small \( \delta \ll k_{\rho } \)
up to second order, we finally get\begin{eqnarray}
\Gamma _{\rho ,\rho }(k) & = & -\frac{\delta ^{2}}{2\pi \hbar m}\left\{ \log \left( \frac{\delta }{8k_{\rho }}\right) +\frac{1}{2}\right\} \nonumber \\
 & \simeq  & -\frac{\delta ^{2}}{2\pi \hbar m}\left\{ \log \left( \frac{\delta }{8k_{F}}\right) +\frac{1}{2}+\rho \gamma \right\} .\label{1} 
\end{eqnarray}
We also expanded in small \( k_{R}\ll k_{F} \) in the second line.

\subsection{Interband case (\protect\( \rho '=-\rho \protect \)).}

We find\begin{equation}
\label{1}
I_{-\rho }=-\frac{2|k-k'|}{kk'}\left[ K(-z)-E(-z)\right] .
\end{equation}
We repeat the same procedure and expand in \( \delta ,k_{R}\ll k_{F} \).
We get\begin{equation}
\label{1}
\Gamma _{\rho ,-\rho }(k)=\frac{\delta ^{2}}{2\pi \hbar m}\left\{ 1+\rho \gamma +\gamma ^{2}\log \frac{\gamma }{4}\right\} .
\end{equation}

We now add the two Rashba branches. The term linear in \( k_{R} \)
vanishes and we finally get for the lifetime including the Rashba
s-o interaction\begin{eqnarray}
\Gamma _{\rho }(k) & = & -\frac{\delta ^{2}}{2\pi \hbar m}\left\{ \log \left( \frac{\delta }{8k_{F}}\right) -\frac{1}{2}-\gamma ^{2}\log \frac{\gamma }{4}  \right\} \label{lifewso} \\
 & = & -\frac{\xi _{k\rho }^{2}}{4\pi \hbar E_{F}}\left\{ \log \left( \frac{\xi _{k\rho }}{16E_{F}}\right) -\frac{1}{2}-\frac{E_{R}}{E_{F}}\log \frac{E_{R}}{8E_{F}}\right\} ,\nonumber \label{1} 
\end{eqnarray}
valid up to  $\delta/(\hbar m) \times O\left( \delta ,\gamma ^{2}\right) $.
We recognize
in the first term the standard lifetime for a 2D Fermi liquid without
Rashba interaction \cite{Giu82,Zhe96}, with the logarithmic enhancement
\( \log (\delta /k_{F})\sim \log (\xi _{k}/E_{F}) \). 

The modification to the lifetime due to the spin-orbit interaction
also contains a logarithmic factor \( \log (k_{R}/k_{F})\sim \log (\Delta _{R}/E_{F}) \)
involving the Rashba splitting at the Fermi surface, \( \Delta _{R}=2\alpha k_{F}/\hbar  \).
We note that for typical GaAs 2DEGs this modification is rather weak,
because of the factor \( \gamma =k_{R}/k_{F}\ll 1 \), and therefore
does not modify significantly the usual term valid without s-o interaction.

\section{Renormalization factor \protect\( Z\protect \)}

We now derive the expression for the renormalization factor \( Z \)
(\ref{zfactor}). We give some details of the calculation, in order
to show the cancellation of the \( \sim \log r_{s} \) term, as well
as to introduce integrations that will also be useful for the calculation
of the effective mass. Our starting point is the real part of the
self-energy entering in Eq. (\ref{zfactor-A}). At \( k_{B}T=0 \),
one can replace \cite{Mah00} the sum over the Matsubara frequencies
appearing in (\ref{selfener}) by an integral along the imaginary
axis, \( k_{B}T\sum _{iq_{n}}f(iq_{n})\to (1/2\pi )\int duf(iu) \).
Thus we need to evaluate\begin{equation}
\label{1}
A=-\frac{\partial }{\partial \xi }\real \sum _{\mathbf{q}\rho '}\frac{1}{2\pi }\int _{-\infty }^{\infty }du\, G_{\rho '}(k',ik_{n}+iu)V'(q,iu),
\end{equation}
 where \( \mathbf{k}'=\mathbf{k}+\mathbf{q} \), \( k=k_{\rho } \)
and one sets \( \xi =0 \) after taking the derivative. While the
analytical continuation \( ik_{n}\to \xi +i0_{+} \) must be taken
after the integration, one can reverse this order (i.e., make the
analytical continuation first), \cite{Mah00} \begin{equation}
\label{ueven}
A=A^{\mathrm{res}}-\sum _{\mathbf{q}\rho '}\frac{\real }{\pi }\int _{0}^{\infty }du\left. \frac{\partial }{\partial \xi }G_{\rho '}(k',\xi +iu)V'(q,iu)\right| _{\xi =0},
\end{equation}
provided that one compensates for the contributions of the poles of
\( G \) by adding the {}``residue'' term \begin{eqnarray}
A^{\mathrm{res}} & = & -\real \sum _{\mathbf{q}\rho '}\frac{\partial }{\partial \xi }\left[ \Theta (-\xi _{k'\rho '})-\Theta (\xi -\xi _{k'\rho '})\right] \nonumber \\
 &  & \hspace {20mm}\times V'(q,\xi _{k'\rho '}-\xi )\Big |_{\xi =0}\nonumber \label{1} \\
 & = & \sum _{\mathbf{q}\rho '}\, \delta (\xi _{k'\rho '})V'(q,0).
\end{eqnarray}
 We have used in Eq. (\ref{ueven}) the fact that the integrated function
is even in \( u \). For the remaining term, we notice that \( -\partial _{\xi }G_{\rho '}(k',\xi +iu)=i\partial _{u}G_{\rho '}(k',iu) \)
when \( \xi =0 \) and integrate by parts over \( u \). The boundary
term with \( u\to \infty  \) vanishes, while the term with \( u\to 0_{+} \)
gives \begin{eqnarray}
A_{\mathrm{boundary}}^{(u\to 0_{+})} & = & \real \, i\frac{1}{\pi }\sum _{\mathbf{q}\rho '}G_{\rho '}(k',i0_{+})V'(q,i0_{+})\nonumber \\
 & = & -\sum _{\mathbf{q}\rho '}\delta (\xi _{k'\rho '})V'(q,0),\label{1} 
\end{eqnarray}
where we have used \( -\imag G_{\rho '}(k',i0_{+})=-\pi \delta (\xi _{k'\rho '}) \).
We see that this boundary term cancel with the residue term \( A^{\mathrm{res}}_{\rho } \).
This is important, as these terms actually contain a term that is
logarithmic in \( r_{s} \) {[}see Eq. (\ref{blog}) in the calculation
of \( m^{*} \) below{]}. Thus we have\begin{eqnarray}
A & = & \imag \frac{1}{\pi }\sum _{\mathbf{q}\rho '}\int _{0}^{\infty }duG_{\rho '}(k',iu)\frac{\partial }{\partial u}V'(q,iu)\label{A-z} \\
 & = & \hspace {-0mm}-\frac{r''}{2\pi ^{2}}\sum _{\rho '}\imag \int _{0}^{\infty }\! \! \! dy''\int _{0}^{2\pi }\! \! \! d\phi \int _{0}^{\infty }\! \! \! dx''f(x'',y'',\phi ).\hspace {8mm}\label{1} 
\end{eqnarray}

We have defined \( r''=m e_0^{2}/k_{\rho}\hbar=k_{\mathrm{TF}}/2k_{\rho } \),
\( y''=mu/qk_{\rho } \), and \( x''=q/2k_{\rho } \). The integrand
is \begin{equation}
\label{1}
f(x'',y'',\phi )=\frac{\mathcal{F}'}{iy''-\mu }\frac{1}{x''\varepsilon ^{2}}\frac{\partial \varepsilon }{\partial y''},
\end{equation}
 where \( \mathcal{F}'(x'',\phi )=1/2+\rho \rho '(1 + 2x''\cos \phi )/2\ell  \)
is the overlap of the eigenspinors, \( \ell (x'',\phi )=\sqrt{1+4x''\cos \phi +4x''^{2}}=k'/k \),
\( \mu (x'',\phi )=\cos \phi +x''+(\rho \rho '\ell -1)\gamma ''/2x'' \)
is the dimensionless energy \( \xi _{k'\rho '} \), and \( \gamma ''=\rho k_{R}/k_{\rho } \)
is a modified s-o strength.

We now consider the RPA limit of high density, which corresponds to
small \( r''\ll1  \). In this case, the dominant contribution comes
from the intraband case (\( \rho '=\rho  \)) with \( x''\ll 1 \),
where we can use the approximations \( \mathcal{F}'\simeq 1+O(x''^{2}) \),
\( x''\varepsilon \simeq x''+r'' a(y')+O(x''^{2}) \), \( \mu \simeq \cos \phi (1+\gamma '')+O(x'') \),
and we have defined \( a(y')=1-y'/\sqrt{y'^{2}+1} \) and \( y'=mu/q\kappa =y''/(1+\gamma '') \).
Defining \( r'=m e_0^{2}/\kappa\hbar =r''/(1+\gamma '') \), we have 
\begin{eqnarray}
A & \simeq  & -\frac{r'}{2\pi ^{2}}\imag \int _{0}^{\infty }dy'\frac{\partial }{\partial y'}a(y')\int _{0}^{2\pi }d\phi \frac{1}{iy'-\cos \phi }\nonumber \\
 &  & \, \, \, \, \, \, \, \, \, \, \, \, \, \, \times \int _{0}^{\infty }dx''\frac{\partial }{\partial x''}\frac{-r''}{x''+ra(y')}\nonumber \label{1} \\
 & = & -\frac{r'}{\pi }\int _{0}^{\infty }dy'\frac{1}{(y^{2}+1)^{2}}\frac{1}{a(y')}\nonumber \\
 & = & -\frac{r'}{\pi }\left( 1+\frac{\pi }{2}\right) +O(r'^{2}).
\end{eqnarray}
The remaining terms (in particular, the contribution from the interband
transition with \( \rho '=-\rho  \)) are neglected as they are \( O\left( r'^{2}\right)  \).
We now use\begin{equation}
\label{1}
r'=r_{s}\left( 1+\gamma ^{2}\right) ^{-1/2}\simeq r_{s}\left( 1-\frac{1}{2}\gamma ^{2}\right) ,
\end{equation}
 and the renormalization factor reads\begin{eqnarray}
Z_{\rho } & = & 1-\frac{r_{s}}{\pi }\left( 1+\frac{\pi }{2}\right) \left( 1-\frac{1}{2}\gamma ^{2}\right) \nonumber \\
 & = & 1-\frac{r_{s}}{\pi }\left( 1+\frac{\pi }{2}\right) \left( 1-\frac{E_{R}}{E_{F}}\right) .\label{zfactor final 2} 
\end{eqnarray}
This result is valid up to \( O(r_{s}^{2},r_{s}\gamma ^{3}) \), so
that the modification from the result (\ref{z0}) without s-o disappears
in the case \( \gamma \ll \sqrt{r_{s}}\ll 1 \). Similarly to the
inverse lifetime, we see that the \( Z \)-factor is independent of
the Rashba band index \( \rho  \), and that its modification appears
only in second order in the strength of the s-o interaction. This
modification can be traced back to the small shift of the Fermi surface
due to the s-o interaction. Without s-o inteaction (\( \gamma =0 \)),
Eq. (\ref{zfactor final 2}) corresponds to the result presented in
Refs. \cite{Bur00,Gal04}.

\section{Effective mass}

The calculation for the effective mass (\ref{mass}) is similar in
spirit to the \( Z \) factor calculation, but is more involved. We
start with\begin{equation}
\label{1}
B=-\frac{m}{\kappa }\frac{\partial }{\partial k}\real \sum _{\mathbf{q}\rho '}\frac{1}{2\pi }\int _{-\infty }^{\infty }du\, G_{\rho '}(k',ik_{n}+iu)V'(q,iu),
\end{equation}
 where \( k\to k_{\rho } \) after taking the derivative. Again, we
first perform the analytical continuation \( ik_{n}\to \xi +i0_{+} \)
by adding a residue term \( \sim \Theta (-\xi _{k'\rho '})-\Theta (\xi -\xi _{k'\rho '}) \).
Contrary to the case for \( Z \), this residue term identically vanishes
once we take \( \xi =0 \). Hence we have 
\begin{equation}
\label{Bintform}
B=-\real \frac{m}{\kappa }\sum _{\mathbf{q}\rho '}\frac{1}{\pi }\int _{0}^{\infty }du\left. \frac{\partial }{\partial k}G_{\rho '}(k',iu)V'(q,iu)\right| _{k=k_{\rho }},
\end{equation}
which we integrate by parts.
With the change of variables $\mathbf{q}\to \mathbf{k}'$, the boundary term reads 
\begin{eqnarray}
B_{\mathrm{boundary}}^{(u\to 0_{+})} 
& = & \frac{m}{\kappa}\sum _{\mathbf{k}'\rho '}\delta (\xi _{k'\rho '})V'(q,0)\frac{\partial \xi _{k'\rho '}}{\partial k} \label{1b1}  \\
 & = &\frac{r'}{8\pi } \sum _{\rho '}\frac{k_{\rho'}}{k_\rho}\int _{0}^{2\pi }d\theta \frac{\cos\theta}{x''_{\rho \rho'}(\theta)+r''}(1+\rho \rho' \cos\theta), \nonumber
\end{eqnarray}
where 
%$x''_{\rho \rho'}(\theta)=q_{\rho \rho'}(\theta)/2k_\rho$ with $q_{\rho \rho'}(\theta)=|\mathbf{k}'-\mathbf{k}|$ for $k'=k_\rho'$, $k=k_\rho$
%$x''_{\rho \rho'}(\theta)= (1/4)\sqrt{ (k_{\rho'}/k_\rho)^2 -2 \cos \theta  k_{\rho'}/ k_{\rho} +  1}$ and 
\protect{ $x''_{\rho \rho'}(\theta)= %q/2k_\rho =
(1/2k_\rho)\sqrt{k^2_{\rho'}-2  k_{\rho'} k_{\rho} \cos \theta +  k^2_{\rho}}$ }
and  we recall that $r'=me_0^2/\kappa \hbar$. 
We consider the case $ r_{s},\gamma \ll 1$ and finally get
\begin{equation}
\label{blog}
B_{\mathrm{boundary}}^{(u\to 0_{+})}=
-\frac{r_s}{\pi }\left[ \log \left( \frac{r_{s}}{2}\right) +2 + \frac{4\rho \gamma}{3}  
-\frac{\gamma^2 }{2} \log\gamma +O(r_s,\gamma^2)\right] .
\end{equation}
The remaining integrated term of the integration by parts in (\ref{Bintform})
contains two terms. The first one reads 
\begin{equation}
\label{1}
B_{\mathrm{int}}^{(1)}=-\imag \frac{m}{\kappa \pi }\sum _{\mathbf{q}\rho '}\int _{0}^{\infty }duG_{\rho '}(k',iu)\frac{\partial }{\partial u}V'(q,iu)\frac{\partial \xi _{k'\rho '}}{\partial k}.
\end{equation}
We see that the integrand is identical to the expression (\ref{A-z})
appearing in the calculation made for \( Z \), apart from the factor
\( \partial \xi _{k'\rho '}/\partial k\simeq (k/m)(1+\rho '\gamma ''+2x\cos \phi ) \)
for \( x\ll 1 \). We proceed as before, and get in lowest order 
\begin{equation}
\label{1}
B_{\mathrm{int}}^{(1)}=\frac{r'}{\pi }\left( 1+\frac{\pi }{2}\right) .
\end{equation}
The second term reads 
\begin{equation}
\label{1}
B_{\mathrm{int}}^{(2)}=-\real \frac{m}{\kappa \pi }\sum _{\mathbf{q}\rho '}\int _{0}^{\infty }duG_{\rho '}(k',iu)
\frac{V_C(q)}{\varepsilon(q,iu)}
\frac{\partial \mathcal{F}'}{\partial k},
\end{equation}
where \( \partial \mathcal{F}'/\partial k=\rho \rho 'q^{2}\sin ^{2}\phi /2k'^{3}. \)
The analysis of this integral is rather demanding, as no approximation
is accurate and only a numerical solution seems possible. However,
a careful examination of the different terms shows that there is no
logarithmic contribution ---in particular, the small \( x \) contributions
are suppressed by the overall \( \sim x^{3} \) dependence.
We can now use the general argument about the self-energy (see Sec. \ref{seflnolin}), which
states that no term linear in $\gamma$ can be present in the effective mass. 
This implies that this integral $B_{\mathrm{int}}^{(2)}$ compensates for the linear term
appearing in (\ref{blog}), which is confirmed by numerical integrations.

Finally, we obtain the effective mass (\ref{mass})
\begin{equation}
\label{m-so}
\frac{m_{\rho }^{*}}{m}-1=\frac{r_{s}}{\pi }
\left( \log r_{s}+2-\log 2 -\frac{1 }{2}\gamma^2 \log\gamma  \right) ,
\end{equation}
where we used the expression (\ref{zfactor final 2}) for \( Z \).
We recognize in the first three terms the unperturbed result
(\ref{meff0}).
The modification induced by the s-o interaction has the form $\gamma^2 \log \gamma$,
similarly to the lifetime, Eq. (\ref{lifewso})
We see that particles in different Rashba spin eigenstates have, to lowest order, the same
effective mass.

\section{Conclusion}

We have calculated the main quasiparticle parameters (the inverse
quasi-particle lifetime \( 1/\tau  \), the renormalization factor
\( Z \), and the effective mass \( m^{*} \)) due to the Coulomb
electron-electron interaction in a 2D Fermi liquid with Rashba interaction.
The modifications due to
s-o interaction are found to be independent of the Rashba band index
\( \rho  \), and to appear only in second order in the s-o strength \( \gamma \sim \sqrt{E_{R}/E_{F}} \)
with some logarithmic enhancement for the lifetime and the effective mass.

The spin-orbit constant being rather small in typical semiconductor 2DEGs, these modification
will be very small, around $10^{-3}$.
For instance, a GaAs 2DEG with \cite{Mil03} 
\( \alpha =0.5\cdot 10^{-12}\, \mathrm{eV}\, \mathrm{m} \),
\( n=4\cdot 10^{15}\, \mathrm{m}^{-2} \) 
yields
\( k_{R}=0.43\, \mu \mathrm{m}^{-1} \).
This gives a rather small \( \gamma =2.7\cdot 10^{-3} \), so that
one gets only very small modifications
Even for an InGaAs 2DEGs with larger s-o coupling and with \cite{Sat01} 
\( \alpha =30\cdot 10^{-12}\, \mathrm{eV}\, \mathrm{m} \)
for \( n=10\cdot 10^{15}\, \mathrm{m}^{-2} \), \( m=0.03m_{e} \)
and \( r_{s}=0.18 \), one has a modest \( \gamma =0.051 \).

We note that replacing the Rashba interaction by the Dresselhaus interaction
\( H_{D}=\beta (-p_{x}\sigma _{x}+p_{y}\sigma _{y})/\hbar  \) yields
exactly the same results. Indeed, the only difference lies in the
eigenspinors (the phase is decreased by \( \pi /2 \)), so that their
overlap \( \mathcal{F}' \), Eq.(\ref{fprime}), is unchanged and
the energies have the same form. Natural extensions of this work are
the studies of the effect of the s-o interaction on the renormalized
\( g \)-factor \cite{Jan69}, the consideration of short-range potential
instead of the Coulomb interaction, finite temperatures, and the presence
of a perpendicular or parallel magnetic field as used to measure
the mobility and the effective mass or to manipulate electron spins.

\begin{acknowledgments}
We thank B. Altshuler, M. Duckheim, J.C. Egues, and S. Erlingsson for useful discussions.
We are particularly grateful to S. Chesi for very valuable remarks.
This work has been supported by NCCR ``Nanoscale
Science'', Swiss NSF, EU-Spintronics, DARPA, ARO, and ONR.
\end{acknowledgments}


\begin{thebibliography}{10}
\bibitem{Lifediffer}Note that the expressions given in Refs. \cite{Giu82} and \cite{Zhe96}
only agree for the dominant logarithm \( \log (\xi /E_{F}) \), while
the rest of \( O(1) \) differ.
\bibitem{rs}The paramater \( \bar{r}_{s}=r_{s}\sqrt{2} \) is also used in the
literature.
\end{thebibliography}
\end{document}